\documentclass[aps,prl,showpacs,twocolumn]{revtex4}

\usepackage{graphicx}

\begin{document}

\title{Cancellation of Spin-Orbit Effects in Quantum Gates Based\\
on the Exchange Coupling in Quantum Dots}

\author{Guido Burkard}
\author{Daniel Loss}
\address{
Department of Physics and Astronomy,
University of Basel,
Klingelbergstrasse 82,
CH-4056 Basel, Switzerland}

\newcommand{\1}{{{\bf S}_1}}
\newcommand{\2}{{{\bf S}_2}}

\newcommand{\bb}{{\mbox{\boldmath $\beta$}}}

\newcommand{\bra}[1]{{\langle #1 |}}
\newcommand{\ket}[1]{{| #1 \rangle}}

\newcommand{\spup}{\ket{\!\uparrow}}
\newcommand{\spdown}{\ket{\!\downarrow}}
\newcommand{\spupup}{\ket{\!\uparrow\uparrow}}
\newcommand{\spupdown}{\ket{\!\uparrow\downarrow}}
\newcommand{\spdownup}{\ket{\!\downarrow\uparrow}}
\newcommand{\spdowndown}{\ket{\!\downarrow\downarrow}}


\begin{abstract}
\hspace{-4mm}
\begin{minipage}{14.9cm}
\hspace{3mm}We study the effect of the spin-orbit interaction
on quantum gate operations based on the spin exchange
coupling where the qubit is represented by
the electron spin in a quantum dot
or a similar nanostructure.
Our main result is the exact cancellation of
the spin-orbit effects in the sequence
producing the quantum {\sc XOR} gate 
for the ideal case where
 the pulse shapes of the exchange 
and spin-orbit interactions are iden\-tical.  For the non-ideal
case, the two pulse shapes can be made almost
identical and
the gate error is strongly suppressed by two small parameters, the
spin-orbit constant
and the deviation of the two pulse shapes.
We show that the dipole-dipole interaction leads only
to very small errors in the {\sc XOR} gate.
\end{minipage}
\end{abstract}



\maketitle


The spin 1/2 of an electron is a
``natural'' representation of a quantum bit (qubit)
since it comprises exactly two levels; there
are no additional degrees of freedom into which the
system could ``leak'' and thereby cause errors in
a quantum computation.
In addition to this, magneto-optical experiments
have revealed unusually long spin coherence times
in doped semiconductors,
exceeding $100\,\mu{\rm s}$ \cite{KA}, thus making
electron spins in semiconductors suitable
candidates for a scalable quantum computer
architecture.
These advantages have motivated the idea of
spin-based solid-state quantum computation using 
electron spins in coupled quantum dots \cite{LD},
where the required
two-spin coupling is provided by the 
Heisenberg exchange interaction between the
two spins in adjacent quantum dots.
The microscopic origin of the exchange coupling
lies in the virtual tunneling of electrons
from one quantum dot to the other and back,
and there are several external physical 
parameters (gate voltages, magnetic field, etc.)
which can in principle be used for 
controlled quantum gate operation \cite{BLD}.
Subsequent schemes \cite{Privman,Kane,Barnes,Vrijen,Levy} for
solid-state quantum computation rely also on the 
exchange interaction between spins,
and it has been pointed out that the exchange
interaction alone (without single-spin manipulation)
is in principle sufficient for universal quantum 
computation \cite{Bacon,DBKBW}.

However, the two-level structure of the {\em spin} of the
electron is only approximate if one includes
relativistic effects which lead to spin-orbit coupling
\cite{footnote-spin}.
The exchange Hamiltonian can acquire anisotropic terms
due to spin-orbit coupling \cite{D,M}.
For conduction band electrons in single GaAs dots, the 
spin-orbit energy
is typically small \cite{BLD}, however it was recently
pointed out by Kavokin \cite{Kavokin} that the
spin-orbit coupling can be relevant for tunneling 
between two dots, leading to an
anisotropy in the resulting spin Hamiltonian,
and it was suggested that it may lead to
additional spin decoherence.
Subsequently, Bonesteel {\em et al.} \cite{BSD} have demonstrated
that the first-order effect of the spin-orbit coupling during
quantum gate operations can be eliminated 
by using time-symmetric pulse shapes for the
coupling between the spins.

In this paper, we present a different method for dealing
with the spin-orbit interaction.  Our main result is
that the spin-orbit effects {\em exactly}
cancel in the gate sequence [Eq.~(\ref{gate})]
required to produce the
quantum {\sc XOR} ({\sc CNOT}) gate, provided that
the pulse form for the spin-orbit and the exchange couplings
are the same.  Since {\sc XOR} is
sufficient  to assemble any quantum computation together
with single-qubit operations, this result
has far-reaching consequences for
spin-based quantum computation with the exchange interaction;
it ascertains that the spin-orbit coupling can be dealt with in
any quantum computation.  In reality, the pulse shapes for the
exchange and the spin-orbit coupling cannot be chosen 
completely identical.  Typically, however, we can
to choose two pulse shapes which are very similar and
show that our result still holds to a very good approximation,
i.e.\ the effect of the spin-orbit coupling is still
strongly suppressed.  Finally, we discuss the effect of
the dipole coupling between adjacent spins, providing
another anisotropic coupling.
The anisotropy due to an inhomogeneous magnetic
field was studied in \cite{Hu}.

The spin-orbit coupling for a conduction-band electron
(momentum ${\bf k}$, spin ${\bf S}$) can be written as
$H_{\rm so} = {\bf h}({\bf k})\cdot {\bf S}$.
In two dimensions, the Rashba term \cite{Rashba}
${\bf h}_1({\bf k}) = a_1 (k_y,-k_x,0)$ arises
from an asymmetric quantum well or from an external field.
The absence of the inversion symmetry,
e.g.\ in GaAs, causes a term \cite{DK}
${\bf h}_2({\bf k}) = a_2 (-k_x,k_y,0)$.
Such a term was already shown to exist in \cite{Altshuler}.
The isotropic Heisenberg coupling with exchange
energy $J$ and the anisotropic
exchange between two localized spins $\1$ and $\2$ ($s=1/2$)
are combined in the Hamiltonian \cite{BSD}
$H(t) = J(t) \left( \1 \cdot \2 +   {\cal A}(t)\right)$.
We divide ${\cal A}(t)$ into asymmetric and symmetric
parts \cite{Kavokin},
\begin{equation}
  \label{a}
  {\cal A}(t) = \bb(t) \cdot (\1\times\2)
           +  \gamma(t) (\bb(t)\cdot\1) (\bb(t)\cdot\2),
\end{equation}
where $\bb = \langle \psi_1 | i {\bf h}({\bf k})|\psi_2\rangle$
is the spin-orbit field,
$|\psi_i\rangle$ the groundstate in dot $i=1,2$,
and $\gamma\approx O(\beta^0)$.
For ${\cal A}=0$, the
quantum {\sc XOR} gate can be obtained by applying $H(t)$
twice, together with single-spin rotations \cite{LD,remark1},
\begin{equation}
  \label{gate}
  U_g = e^{i \pi S_1^z/2}e^{-i \pi S_2^z/2}\, U \, e^{i\pi S_2^z} \, U\, ,
\end{equation}
where $U$ is the (unitary) time-ordered exponential
$U = T\exp(-i \int_{-\tau_s/2}^{\tau_s/2} \! H(t) \, dt )$.
Here $\tau_s$ denotes the switching time, during
which the spin interactions via tunneling are turned on.
In the case ${\cal A}=0$, the Hamiltonian commutes
with itself at different times and thus $U$ is only a 
function of the integrated interaction strength,
\begin{equation}
  \label{phi}
  \varphi = \int_{-\tau_s/2}^{\tau_s/2}\!J(t)\,dt \, ,
\end{equation}
with $\varphi\neq 0$.
In particular, we obtain the desired quantum gate
(up to a trivial change of basis)
$U_g = U_{\sc CPF} = e^{i\pi S_2^y/2}\, U_{\sc XOR} \, e^{-i\pi S_2^y/2}$
if we choose $\varphi=\pi/2$ (in this case, $U$ is the
``square-root of swap'' gate \cite{LD}).

First, we study the case ${\cal A} \neq 0$, retaining
the property that $H(t)$ commutes with itself at different
times.  This is the case if $\bb$ and $\gamma$ (and thus ${\cal A}$) 
are time-independent,
i.e.\ if the anisotropic part of the Hamiltonian $H$
is proportional to the isotropic exchange term. 
This allows us to fix a coordinate
system in which $\bb$ points along the $z$ axis, and in which
the anisotropy can be written as
\begin{equation}
  \label{a0}
  {\cal A} = \beta (S_1^x S_2^y - S_1^y S_2^x) + \delta S_1^z S_2^z ,
\end{equation}
with $\delta=\gamma\beta^2$.
In this basis $H$ commutes with the $z$ component $S^z = S_1^z + S_2^z$ of
the total spin, $[ H, S^z]=0$, and thus $\spupup$ and
$\spdowndown$, being non-degenerate eigenstates of $S^z$, are also
eigenstates of $H$. 
Note that in their energy eigenvalue $J(1+\delta)/4$ there is
no contribution from the first term in Eq.~(\ref{a0}).
In the $S^z=0$ subspace we choose a basis consisting of
the spin singlet
$\ket{s} = (\spupdown - \spdownup)/\sqrt{2}$ and the triplet
$\ket{t} = (\spupdown + \spdownup)/\sqrt{2}$
because this choice makes the
isotropic part $J \1\cdot\2$ of the Hamiltonian diagonal.
The complete Hamiltonian in the basis
$\{\spupup, \ket{s}, \ket{t}, \spdowndown\}$ is 
\begin{equation}
  \label{hmatrix}
  H(t) = \frac{J(t)}{2}\left(\begin{array}{c c c c}
1+\delta & 0       & 0      & 0   \\
0   & -1      & i\beta & 0   \\
0   & -i\beta & 1      & 0   \\
0   & 0      &  0      & 1+\delta    
\end{array} \right),
\end{equation}
where we have added an irrelevant term $J(1+\delta)/4$
proportional to the unity matrix.

Exponentiation of Eq.~(\ref{hmatrix})
in the  $S^z=0$ subspace yields
\begin{equation}
  \label{sa0}
  U\Big|_{S^z=0} = \left(\begin{array}{c c}
c+is/x & \beta s \\
- \beta s &  c-is/x
\end{array} \right),
\end{equation}
where $c=\cos(x \varphi /2)$, $s=\sin(x \varphi /2)$,
$x=\sqrt{1+\beta^2}$ and where $\varphi$ is defined in Eq.~(\ref{phi}).
Since $\exp(i\pi S_2^z)=-i\sigma_x$ in the
$S^z=0$ subspace, we find
$U e^{i\pi S_2^z} U|_{S^z=0} = - i \sigma_x$,
i.e.\ the dependence of $U|_{S^z=0}$ on the phase $\varphi$ as well as
on the spin-orbit parameter
$\beta$ as shown in Eq.~(\ref{sa0}) cancels exactly in the sequence
Eq.~(\ref{gate}).
In other words, when we construct the {\sc XOR} gate
there will be no effect of the
time-independent anisotropic terms ${\cal A}$ in the $S^z=0$ subspace.
By a proper choice of $\varphi$,
we can also eliminate the effect of the anisotropy for the
states $\spupup$ and $\spdowndown$.  This can be seen by writing
down the full unitary operator Eq.~(\ref{gate}) using the Hamiltonian
Eq.~(\ref{hmatrix}),
\begin{equation}
  \label{umatrix}
  U_g={\rm diag}(ie^{-i\varphi(1+\delta)},1,1,-ie^{-i\varphi(1+\delta)}),
\end{equation}
where ${\rm diag}(x_1,\ldots , x_4)$ denotes the diagonal matrix with
diagonal entries $x_1,\ldots , x_4$.
The pulse strength $\varphi$ and the spin-orbit parameters only
enter $U$ in the $S^z=\pm 1$ subspaces.
We would like $U_g$ to be the conditional phase flip operation
$U_{\sc CPF}={\rm diag}(1,1,1,-1)$, being equivalent to the {\sc XOR}
operation up to a basis change.
Indeed, the condition $U_g=U_{\sc CPF}$ can be fulfilled
for $\varphi=\pi/2(1+\delta)$.

We have shown that in the case where the anisotropic term
in the Hamiltonian is proportional to the isotropic term
(i.e.\ ${\cal A}={\rm const.}$),
we can completely eliminate the effect of the anisotropy
by a proper choice of the pulse strength $\varphi$.
In real systems, however, the anisotropic terms in the 
Hamiltonian $H$ cannot be expected to be 
exactly proportional to $J(t)$, i.e.\ ${\cal A}(t)$
is time-dependent.  In general, both $\bb$ and $\gamma$
depend on time.  Under these circumstances, we cannot exactly
eliminate the effect of the anisotropy 
because of the time-ordering in the definition of $U$ and
since the Hamiltonian does, in general,
not commute with itself at different times, $[H(t),H(t')]\neq 0$.

In the following, we estimate the errors due
to the anisotropy in the Hamiltonian in the case where
${\cal A}$(t) is only weakly time-dependent.  Subsequently,
we present a procedure which allows us to achieve exactly this
situation (i.e.\ a weakly time-dependent ${\cal A}$).
We write ${\cal A}(t)={\cal A}_0+\Delta{\cal A}(t)$,
where ${\cal A}_0$ is constant, as in Eq.~(\ref{a0}),
and $\Delta{\cal A}(t)$ is the small
time-dependent deviation from ${\cal A}_0$.
The Hamiltonian is written as the sum
$H(t)=H_0(t) + H'(t)$ where $H_0(t)$ is given by Eq.~(\ref{hmatrix})
and 
\begin{eqnarray}
   H'(t) &=& J(t)\Delta{\cal A}(t) \label{dev}\\
         &=& J(t)\left(\Delta\bb(t) \cdot (\1\times\2)
           +  \Delta\gamma(t) (\bb \cdot\1) (\bb\cdot\2) \right).
            \nonumber
\end{eqnarray}

Note that in the symmetric part
we have already omitted terms which are of order $\Delta\beta \Delta\gamma$.
This Hamiltonian generates a unitary time evolution $U = U_0 + \Delta U$,
where $U_0=T \exp (-i\int_{-\tau_s/2}^{\tau_s/2}H_0(t)\,dt)$ is the contribution
due to $H_0$.
The explicit form of $\Delta U$ is rather complicated;
however, we are only interested
in estimating the gate error $\Delta U_g = U_g - U_{\sc CPF}$
caused by $H'(t)$ (note that $\Delta U_g$ is not unitary).
For this purpose we work in the interaction picture
with respect to $H_0(t)$,
where $U_I=U_0^\dagger U  =  T\exp(-i\int_{-\tau_s/2}^{\tau_s/2}H_I'(t)\,dt)$
and $H'_I=U_0^\dagger  H' U_0$. In this representation, the
deviation $\Delta U$ from the ``ideal'' time evolution $U$ becomes
$\Delta U_I = U_I - 1 = -i\int_{-\tau_s/2}^{\tau_s/2}H_I'(t)\,dt + O(H_I'^2)$.
The norm of the gate error
$||\Delta U_g||=\max_{\langle\psi|\psi\rangle=1}
\sqrt{\langle\psi|\Delta U_g^\dagger\Delta U_g|\psi\rangle}$
(to lowest order in $H'$) can now be estimated as follows,
\begin{eqnarray}
\label{norm}
  ||\Delta U_g|| & \lesssim &   2\,  ||\Delta U|| 
                              = 2\,  ||U_0 \Delta U_I|| 
                              = 2\,  ||\Delta U_I||      \nonumber\\
                 & \lesssim & 2\, \tau_s \max_{|t|\le \tau_s/2} ||H'(t)||
                   \equiv 2 \Delta\, ,  
\end{eqnarray}
where the first equality comes from Eq.~(\ref{gate}) and the unitarity of the
involved quantum gates.
Using $\beta\ll 1$, we approximate
$||H'(t)|| \lesssim |J(t)\Delta\bb(t)|/2$, since the second term in
$H'$ is $O(\beta^2)$.
We use $\Delta\beta(t) = \beta(t)-\beta_0$ to write
\begin{equation}
\label{delta}
\Delta  = \frac{|\varphi | \beta_0}{2} \max_{|t|\le \tau_s/2} 
          \left|\frac{J(t)}{J_0}\left(\frac{\beta(t)}{\beta_0}-1\right)\right|, 
\end{equation}
where $J_0$ denotes the average exchange coupling,
$J_0 = \varphi/\tau_s\neq 0$.
Note that
the position of the dots is fixed
during the switching process, thus
$\bb(t)/|\bb(t)|={\rm const.}$ \cite{Kavokin}.
For the {\sc XOR} gate, $\varphi\approx \pi/2$.
The error probability for the described gate operation can now
be estimated as
$\epsilon \equiv ||\Delta\Psi_{\rm out}||^2 \le ||\Delta U_g\Psi_{\rm in}||^2 
               \le  ||\Delta U_g||^2 \lesssim 4 \Delta^2$.

\begin{figure}[t]
\centerline{
\includegraphics[width=5.2cm]{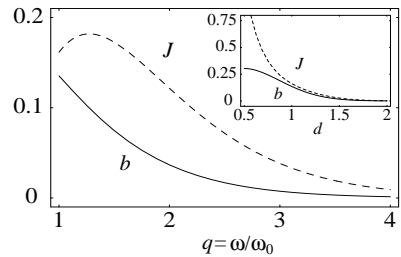}}
\vspace{-2mm}
\caption{\label{Jbplot}
The exchange coupling $J$ (dashed line) in units of $\hbar \omega_0$
and the spin-orbit field $b=\beta J$ (solid line) in units of $b_0$ 
for two electron
spins located in adjacent quantum dots as a function of the
dimensionless parameter $q=\omega/\omega_0$ at fixed interdot
distance $d=1$
(Inset: as a function of $d$ at fixed $q=1$),
where $\hbar\omega$ is the
single-dot confinement energy, $\omega_0$ is the 
(fixed) minimum value of $\omega$, and  $b_0$ is the
spin-orbit parameter.
For this plot, 
$\hbar\omega_0 = 6\,{\rm meV}$ and $c=1.71$.}
\vspace{-5mm}
\end{figure}

In order to obtain an estimate for $\Delta$,
we consider the case of coupled quantum dots in a 2DEG.
For parabolic confinement potential
$V(r)=m \omega r^2/2$, the ground-state orbitals are
$\psi(r)=(\pi a_B^2)^{-1}\exp(-r^2/2 a_B^2)$, where $a_B=\sqrt{\hbar/m\omega}$
is the effective Bohr radius of the electronic orbitals and
$m$ is the effective electron mass.  If two such quantum dots
containing one electron each are separated by a distance $2a$, the
exchange coupling between the spins of the electrons at zero magnetic
field is given by \cite{BLD}
\begin{equation}
  \label{j}
  J(d,q) = \frac{\hbar\omega_0}{\sinh(2qd^2)}\left(c (e^{-qd^2}I_0(qd^2)-1)
                                     +\frac{3q}{4}(1+qd^2)\right),
\end{equation}
where $I_0$ is the zeroth order Bessel function,
$d=a/a_B^0$ the dimensionless ratio between the
half-distance $a$ and the effective
Bohr radius $a_B^0=\sqrt{\hbar/m\omega_0}$,
$c$ characterizes the strength of the (bare) Coulomb interaction
($\hbar\omega_{0}=6\,{\rm meV}$ and $c=1.71$ in our numerical example),
and $q=\omega/\omega_0$ is the strength of the confinement $\omega$
in units of its minimum value $\omega_0$.
Following \cite{Kavokin}, we find for both ${\bf h}_1$ and
${\bf h}_2$ that
$b(d,q) \equiv |J(d,q) \bb(d,q)|=b_0  \sqrt{q} d \exp(-2 q d^2)$,
where $b_0=a_i/a_B^0$, $i=1,2$.
For ${\bf h}_2$ in a $5\:{\rm nm}$ wide [100] GaAs quantum well
$a_2 \approx 2\:{\rm meV}\,{\rm nm}$, or
 $\beta\approx 0.02$ at $d=q=1$.
In Fig.~\ref{Jbplot}, we plot $J(d,q)$ and $b(d,q)$.

\begin{figure}[b]
\vspace{-4mm}
\centerline{
\includegraphics[width=5.1cm]{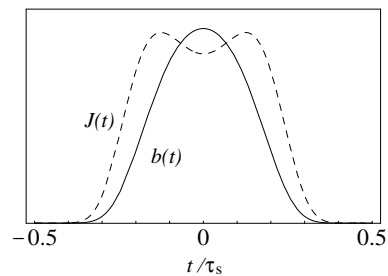}}
\vspace{-2mm}
\caption{
Pulse form of the exchange coupling $J(t)$ (dashed line) and the
spin-orbit field $b(t)=J(t)\beta(t)$ (solid line) for a simple
model involving two coupled quantum
dots which are coupled and decoupled with the time-dependent
confinement strength $q(t)=\omega(t)/\omega_0$.
For the distance between the dots we choose $d=1$.
The choice of the vertical scaling of the two pulses in this graph
is such that the deviations of one pulse from the other are
(approximately) minimal.}
\label{pulses}
\end{figure}

The switching process can be modeled e.g.\ by a
time-dependent distance $d$ between the dots or by
a time-dependent confinement strength $q$.
Here, we choose the latter possibility and use a pulse
$q(t) = \omega(t)/\omega_0 = \cosh ^2 (\alpha t/\tau_s )$,
where we choose $\alpha=4$.
This pulse shape is suited for adiabatic switching
\cite{BLD,Schliemann}
and leads to a pulsed exchange interaction $J(t)=J(d,q(t))$
and spin-orbit field $b(t)=b(d,q(t))$,
where $-\tau_s/2\le t\le \tau_s/2$.
The pulse shapes of the resulting exchange
coupling $J(t)$ and spin-orbit field $b(t)$ are plotted in
Fig.~\ref{pulses}.  In our example, the switching time
amounts to $\tau_s = \pi/2J_0 \approx 140\:{\rm ps}$.
Note that a pulsed switching by electrostatic lowering of the
tunneling barrier between the dots or by applying a magnetic
field results in a very similar time-dependence of $J$ and $b$
and a similar analysis could also be done in these cases.
If the pulse shapes of $b$ and $J$ were identical,
then the effect of the spin-orbit coupling
in the {\sc XOR} gate could be eliminated exactly
(as explained above).

The optimal choice of $\beta_0$ (i.e.\ the one which
minimizes $\Delta$) in our numerical example 
turns out to be $\beta_0 \simeq \beta(t=0.1 \tau_s)$ and
from Eq.~(\ref{delta}) we find $\Delta\approx  7\cdot 10^{-3}$.
Therefore, the gate errors occur at a rate 
$\epsilon \lesssim  4 \Delta^2 \approx  2\cdot 10^{-4}$ 
which is around the currently known threshold
for fault tolerant quantum computation \cite{Preskill}
and could therefore be corrected by quantum error correction.
  Note that in cases where the error $\epsilon$ is
too large for quantum error correction, it can be further
reduced at the cost of a slower gate operation.
This can be achieved by
designing pulses with smaller intensity, where there is
a long period of constant ${\cal A}$ between the rise
and fall of the pulse.

We finally include the dipole-dipole interaction
$H_d = \eta \left( \1\cdot\2 - 3(\1\cdot{\bf \hat{a}})\right)$,
being another source for anisotropic coupling among the
spins $\1$ and $\2$, into our discussion.
Here ${\bf \hat{a}}$ denotes the unit vector pointing
from the center of one to that of the other dot.
The coupling parameter $\eta = \mu_0 g^2 \mu_B^2/4\pi (2a)^3$ is
typically much smaller than the spin-orbit energy,
for $g=2$ and $a=20\:{\rm nm}$ we obtain
$\eta \approx 3\cdot 10^{-12}\:{\rm eV}$, corresponding
to a dipole field of $B_d=\eta/\mu_B\approx 0.5 \:{\rm mG}$.
Nevertheless, we show here that in cases
where the dipole interaction matters (e.g.\ if $g$
is very large), it can again be
dealt with by using the methods described above.
An essential difference between the dipole
and the spin-orbit interactions is that the
dipole interaction between two spins located
in adjacent quantum dots with fixed distance
cannot be changed by applying gate voltages
or static magnetic fields and therefore remains
constant during the entire switching process
and in the ``idle'' time between the switching.
In the following, we assume for simplicity 
that ${\cal A}$ is independent of time, as in
the first part of our discussion.
In addition to this, we assume for the
moment that $J$ is also constant.

When investigating the combined effect of 
the spin-orbit and dipole couplings, we
will restrict ourselves to the
two cases ${\bf h}_1$ and ${\bf h}_2$.
For ${\bf h}_2$, the spin-orbit field is
parallel to the interdot coupling direction
$\bb \parallel {\bf \hat{a}}$ and thus the second term
in $H_d$ has the same form as
the symmetric term in Eq.~(\ref{a0}). 
We set $\xi=\eta/J$ and find 
\begin{widetext}
\vspace{-7mm}
\begin{equation}
  \label{H-parallel}
  H=\frac{J}{2}\left(\begin{array}{c c c c}
   1 + \delta  - 2 \xi  & 0 & 0 & 0 \\
   0 & -1 - \xi  & i \beta  & 0 \\
   0 & -i \beta  & 1 + \xi  & 0 \\
   0 & 0 & 0 & 1 + \delta  - 2\xi   
  \end{array} \right).
\vspace{-2mm}
\end{equation}
Using Eqs.~(\ref{gate}) and (\ref{phi}) with
$\varphi=\pi/2(1+\delta -2\xi)$, we exactly obtain
$U_g=U_{\sc CPF}$, i.e., the combined effect of
the spin-orbit and dipole coupling is eliminated.
For ${\bf h}_1$, we find $\bb\perp {\bf \hat{a}}$.
Choosing ${\bf \hat{a}}$ along the $x$-axis we obtain
\vspace{-2mm}
\begin{equation}
  \label{H-perp}
  H = \frac{J}{2}\left(\begin{array}{c c c c}
    1 \!+\! \delta  \!+\! \xi  & 0 & 0 & -3\xi/2 \\
    0 & -1 + \xi/2 & i \beta  & 0 \\
    0 & -i\beta  & 1 - \xi/2 & 0 \\
    -3 \xi/2 & 0 & 0 & 1 \!+\! \delta \!+\! \xi   
  \end{array} \right).
\vspace{-2mm}
\end{equation}
\end{widetext}
Setting $\varphi=\pi/2(1+\delta +\xi)$, we obtain again
$U_g=U_{\sc CPF}$, therefore
it is possible to eliminate the spin-orbit
and dipole coupling effects  also in this case.

In principle, the analysis for time-dependent
exchange and spin-orbit coupling can be 
repeated including the dipole interaction.
However, the dipole interaction cannot easily be switched
on and off, and therefore, the ``pulse shape''
of the dipole interaction is a constant, i.e.\
very different from those of the exchange and
spin-orbit couplings.
Nevertheless, since the dipole interaction is
usually very small, we can still use 
Eq.~(\ref{norm}) to obtain a reasonable upper
bound on the error by setting $H' = H_d$.
We obtain
$\Delta_d = \tau_s\eta = |\varphi| \eta/J_0$
which for typical numbers (as above,
$J_0\approx {\rm meV}$)
is tiny, $\Delta_d \approx 10^{-9}$.
The error $\epsilon_d = 4 \Delta_d^2 = 4 (\tau_s \eta)^2$ caused
by the dipole interaction is therefore
negligible in typical situations, and
we only have to take it into account
if for some reason (e.g.\ large $g$)
the dipole interaction becomes 
unusually large.

We conclude that while the spin-orbit 
interaction can cause weak decoherence in the
combination with phonons \cite{Khaetskii},
its direct effect on quantum gate operations
{\em nearly cancels} if the pulse shapes of the
exchange and spin-orbit couplings are 
as similar as possible.  We have shown that
in a typical case involving two tunnel-coupled
quantum dots this is easily achievable.
A simple estimate shows that the dipole
interaction between the spins 
is usually much smaller than the spin-orbit 
interaction and can be neglected.  
Nevertheless, we have shown that in cases
where the dipole effects are unusually 
large, the combined effect of spin-orbit and
dipole coupling can be corrected.

We acknowledge 
support from the Swiss NSF and from 
the DARPA QUIST and SPINS programs.



\begin{references}
\bibitem{footnote-spin}
Due to spin-orbit coupling, the Kramers doublet consists of states
in which spin and charge are correlated.
\bibitem{remark1}
We choose a sequence
which is favorable for $J>0$ and which is slightly different
from the one in \cite{LD}.
\end{references}
\end{document}